\documentstyle[aps,prl,preprint,floats,epsfig]{revtex}

\begin{document}

\preprint{\tighten\vbox{\hbox{\hfil CLNS 00/1687}
                        \hbox{\hfil CLEO 00-16}
}}

\title{Study of $\tau$ Decays to Six Pions and Neutrino}  

\author{CLEO Collaboration}
\date{\today}

\maketitle
\tighten

\begin{abstract} 
The $\tau$ decays to six-pion final states have been
studied with the CLEO detector at the Cornell Electron Storage Ring.
The measured branching fractions are 
${\cal B}(\tau^-\to 2\pi^-\pi^+3\pi^0\nu_{\tau})=(2.2 \pm 0.3 \pm 0.4)\times 10^{-4}$
and 
${\cal B}(\tau^-\to 3\pi^-2\pi^+\pi^0\nu_{\tau})=(1.7 \pm 0.2 \pm 0.2)\times 10^{-4}$.
A search for substructure in these decays shows that they are saturated
by intermediate states with $\eta$ or $\omega$ mesons.
We present the first observation of the decay
$\tau^-\to 2\pi^-\pi^+\omega\nu_{\tau}$ and the branching fraction is
measured to be
$(1.2 \pm 0.2 \pm 0.1) \times 10^{-4}$. 
The measured branching fractions are in good agreement
with the isospin expectations
but somewhat below the Conserved-Vector-Current predictions.

\end{abstract}
\newpage

{
\renewcommand{\thefootnote}{\fnsymbol{footnote}}

\begin{center}
A.~Anastassov,$^{1}$ J.~E.~Duboscq,$^{1}$ E.~Eckhart,$^{1}$
K.~K.~Gan,$^{1}$ C.~Gwon,$^{1}$ T.~Hart,$^{1}$
K.~Honscheid,$^{1}$ D.~Hufnagel,$^{1}$ H.~Kagan,$^{1}$
R.~Kass,$^{1}$
% ...
\end{center}
 
\small
\begin{center}
$^{1}${Ohio State University, Columbus, Ohio 43210}\\
% ...
\end{center}

\setcounter{footnote}{0}
}
\newpage

The decays of the $\tau$ lepton provide a good test
of the Standard Model predictions for the hadronic weak current.
The six-pion branching fractions are related to the isovector part 
of the $e^+e^-$ annihilation cross section by the Conserved Vector 
Current (CVC) hypothesis. 
Isospin symmetry relates the relative branching fractions of the decays
$\tau^- \to 2\pi^-\pi^+3\pi^0 \nu_\tau$,
$\tau^- \to 3\pi^-2\pi^+\pi^0\nu_\tau$, and
$\tau^- \to \pi^-5\pi^0\nu_\tau$.
Therefore, the study of six-pion decays
can be used to test the CVC hypothesis and isospin predictions.
A better understanding of the resonance substructure in the
decays is also of particular interest because of the potential
application in suppressing the hadronic background in the
measurement of the $\tau$ neutrino mass.
In this Letter, we present a study of the decays~\cite{charge}
$\tau^- \to 2\pi^-\pi^+3\pi^0 \nu_\tau$ and
$\tau^- \to 3\pi^-2\pi^+\pi^0\nu_\tau$.
This includes measurements of the branching fractions and
the search for resonance substructure.
The latter results are used to
identify the vector and axial-vector current contributions
to the inclusive branching fractions and allow the
proper comparison of the results with the CVC and isospin symmetry
predictions.

The data used in this analysis were collected with the CLEO
detector~\cite{CLEO_detector} at the Cornell Electron Storage Ring
(CESR) at center-of-mass energy of $10.6$~GeV.
The sample corresponds to a total integrated luminosity
$13.5~\rm fb^{-1}$ and contains
$12.3 \times 10^{6}~\tau^+\tau^-$ events~\cite{KORALB}.

For the decay  $\tau^- \to 2\pi^-\pi^+3\pi^0 \nu_\tau$ 
($3\pi^-2\pi^+\pi^0\nu_\tau$)
we select events with four (six) charged tracks and zero
net charge. The momentum of each track must be greater than 
$100$~MeV and the polar angle $\theta$ of each track with respect to the
beam must satisfy $|\cos\theta|<0.90$.
The track must be consistent with originating from the $e^+e^-$
interaction point. 
This vertex requirement also suppresses the background
from $\tau$ events with a $K_S$ or photon conversion.
The $K_S$ background is further reduced by rejecting events
containing a detached vertex with a $\pi^+\pi^-$ mass consistent with
the nominal $K_S$ mass.

We define two exclusive sets of photon candidates in the barrel
calorimeter ($|\cos\theta| < 0.80$):
high quality (HQ) and low quality (LQ) photons. 
The selection criteria for HQ photons are designed to
minimize the contamination of fake showers from hadronic
interactions, while the acceptance of 
LQ photons ensures high event detection efficiency.
A HQ photon must have an energy $E_\gamma >120$~MeV and a lateral 
profile of energy deposition consistent with that expected for a photon.
The Fisher discriminant technique~\cite{fisher} is used
to differentiate a real photon from a fake photon.
The discriminant function is a linear combination of the
energy of the shower and its distance to the intersection of the
nearest charged track with the calorimeter surface.
We use Monte Carlo simulated events (see below)
to obtain the discriminant function
that maximizes the separation between real and fake photons.
A LQ photon is defined as a shower that does not pass the HQ photon 
requirements but has $E_\gamma > 30$~MeV.
The Fisher discriminant for differentiating a real photon from a fake
photon is only used in the decay $\tau^- \to 2\pi^-\pi^+3\pi^0\nu_{\tau}$
due to the higher photon multiplicity.
HQ photons in the endcap calorimeter ($0.80<|\cos \theta |<0.95$)
are selected as in the barrel.

Each event is divided into two hemispheres using the plane
perpendicular to the thrust axis~\cite{thrust}, calculated using both
charged tracks and photons.
There must be one charged track in one hemisphere (tag) recoiling
against three or five charged tracks in the other hemisphere (signal),
depending on the decay mode. 

In the tag hemisphere, the total invariant mass of charged
tracks and photons must satisfy $M_{tag} < 0.5 \rm~GeV$.
In the signal hemisphere, there should be at least six (two) photons
forming three (one) $\pi^0$ candidates for the decay  
$\tau^- \to 2\pi^-\pi^+3\pi^0 \nu_\tau$ ($3\pi^-2\pi^+\pi^0\nu_\tau$).
The $\pi^0$ candidates are reconstructed from photon
pairs in the barrel calorimeter.
All HQ photons in the barrel must be used in the reconstruction
and no HQ photon in the endcap is allowed.
In case of multiple entries, we select the photon combination with
smallest $\chi^2 = \sum_{i=1}^{3(1)} [S_{\gamma \gamma}^2]_i$,
where $S_{\gamma\gamma} = 
(m_{\gamma \gamma}-m_{\pi^0})/\sigma_{\gamma\gamma}~$
($\sigma_{\gamma\gamma}$ is the mass resolution calculated
from the energy and angular resolution of each photon).
The signal region for the $\pi^0$ candidates is defined
as $-3.5<S_{\gamma\gamma}<2.5$.
In the case of $\tau^- \to 3\pi^-2\pi^+\pi^0\nu_\tau$
we use sideband subtraction to estimate the fake $\pi^0$ background,
with sidebands defined as
$-8.5<S_{\gamma\gamma}<-5.5$ and $4.5<S_{\gamma\gamma}<7.5$.
The total invariant mass of the hadronic system in the signal 
hemisphere must satisfy $M_{6\pi} < M_\tau = 1.777 \rm\ GeV$~\cite{PDG98}.  
The signal hemisphere must have a positive pseudo-neutrino mass:
$$M^{2} = {M_\tau^2 + M_{6\pi}^2 -2M_\tau E^*_{6\pi}} > 0\quad,$$
where $E^*_{6\pi}$ is the energy of
the six-pion system in the $\tau$ rest frame, assuming that the $\tau$
has the full beam energy and that the
$\tau$ direction is given by the momentum vector of the six-pion system.
This cut selects events with tau-like kinematics,
suppressing both the hadronic background and the $\tau$ migration
background from lower multiplicity decays where the six-pion
momentum is not as good of an approximation of the $\tau$ direction.
The background from $\tau$ decays with Dalitz $\pi^0$ decays
or photon conversion in inner detector material
is suppressed by requiring that any pairs of oppositely charged
tracks have invariant mass exceeding 120 MeV when
at least one of the pair is identified as an electron
and assigning the electron mass to both tracks. The
electron identification requires that a track is matched
to a shower with shower energy to momentum ratio exceeding 0.85,
and, if available, specific ionization in the drift chamber
consistent with that expected for an electron.
Two-photon backgrounds are eliminated by requiring the direction of the
missing momentum of the event to satisfy $|\cos \theta_{missing}| < 0.9$.

\begin{figure}
\centering
\centerline{\hbox{\psfig{figure=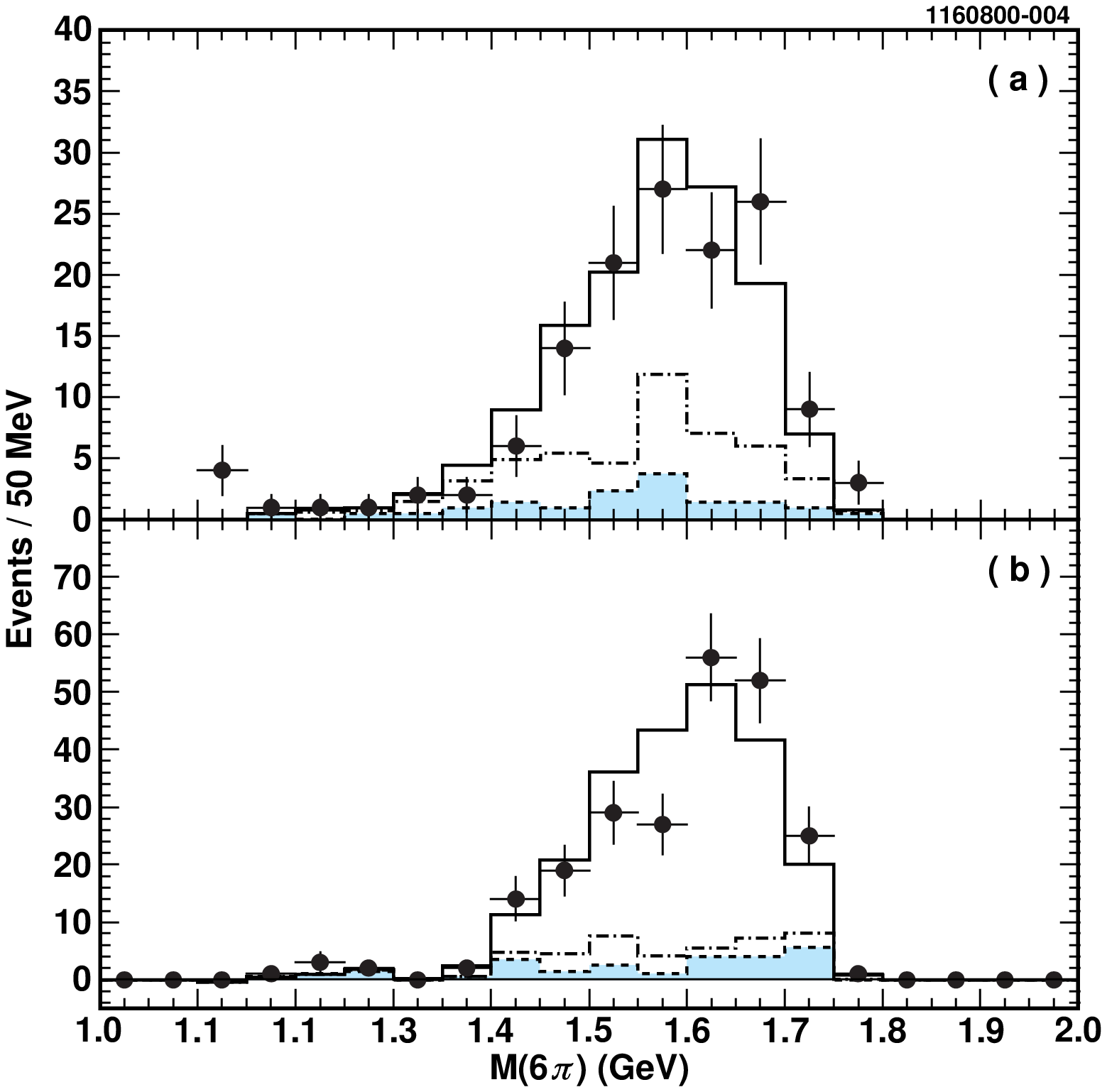, width=8.0cm}}}
\vspace{0.2in}
\caption{Mass spectra of the hadronic systems in
         the decays (a) $\tau^-\to 2\pi^- \pi^+ 3\pi^0 \nu_{\tau}$
         and (b) $\tau^-\to 3\pi^- 2\pi^+ \pi^0 \nu_{\tau}$.
         The solid histogram is the sum of the signal Monte Carlo and
         background (dashed), which includes the $\tau$ migration and hadronic
         (shaded) background.}
\label{figure:mass_spectra}
\end{figure}

The detection efficiency and $\tau$ migration background
are calculated with a Monte Carlo technique.
We use the KORALB/TAUOLA program~\cite{KORALB} for the $\tau$ event simulation.
The decay $\tau^- \to 2\pi^-\pi^+3\pi^0\nu_\tau$ is modeled using a mixture of
$\tau^- \to 2\pi^-\pi^+\eta\nu_\tau$, $\tau^- \to \pi^-2\pi^0\eta\nu_\tau$,
and $\tau^- \to \pi^-2\pi^0\omega\nu_\tau$.
The other decay $\tau^- \to 3\pi^-2\pi^+\pi^0\nu_\tau$ is modeled using a mixture of
$\tau^- \to 2\pi^-\pi^+\eta\nu_\tau$ and $\tau^- \to 2\pi^-\pi^+\omega\nu_\tau$.
The relative mixtures are determined from the measured branching
fractions presented in this and a previous Letter~\cite{3pieta}.
We assume that the $3\pi\eta$ decays proceed through $\pi f_1$ with a spectral function
dominated by the form factor of the $a_1(1260)$ resonance~\cite{Li}.
The $3\pi\omega$ system is modeled assuming dominance of
the $\rho(1700)$ resonance.
The detector response is simulated using the GEANT program~\cite{GEANT}. 
The hadronic background is calculated empirically using a sample of
high-mass tagged events ($1.8 < M_{tag} < 2.8 \rm~GeV$) obtained
from the data with the assumption that $M_{tag}$ and $M_{6\pi}$ are not correlated.
The invariant mass spectra of the six-pion hadronic systems after
applying all selection criteria are shown in Fig.~\ref{figure:mass_spectra}.
There is good agreement between the data and expectation.
The signal, background, and detection efficiency are summarized
in Table~\ref{table:6pi_summary}.
The $\tau$ background includes the decays with $K_S \to \pi^+\pi^-$
and $\pi^0\pi^0$.

\begin{table}
\caption{Summary of the results for $\tau^-\to 2\pi^- \pi^+ 3\pi^0 \nu_{\tau}$
         and $\tau^-\to 3\pi^- 2\pi^+ \pi^0 \nu_{\tau}$.
         All errors are statistical, except the second errors in the branching
         fractions, which are systematic.}
\begin{center}
\begin{tabular}{lcc} 
Decay mode   &$2\pi^- \pi^+ 3\pi^0 \nu_{\tau}$ & $3\pi^- 2\pi^+ \pi^0 \nu_{\tau}$ \\
\hline 
Data (events)            &  $139.0\pm 11.8$          &  $231.0 \pm 18.8$ \\
$q \bar q$ bg (events)   &  $15.1 \pm 3.1$           &  $25.8  \pm 5.9 $ \\
$\tau$ bg (events)       &  $35.2 \pm 3.4$           &  $19.4  \pm 5.5 $ \\
Efficiency (\%)          &  $1.65 \pm 0.03$          &  $4.45  \pm 0.06$ \\ \hline
${\cal B}~(10^{-4})$     &  $2.2 \pm 0.3 \pm 0.4$    &  $1.7   \pm 0.2 \pm 0.2$ \\ 
\end{tabular}
\end{center}
\label{table:6pi_summary}
\end{table}

\begin{figure}
\centering
\centerline{\hbox{\psfig{figure=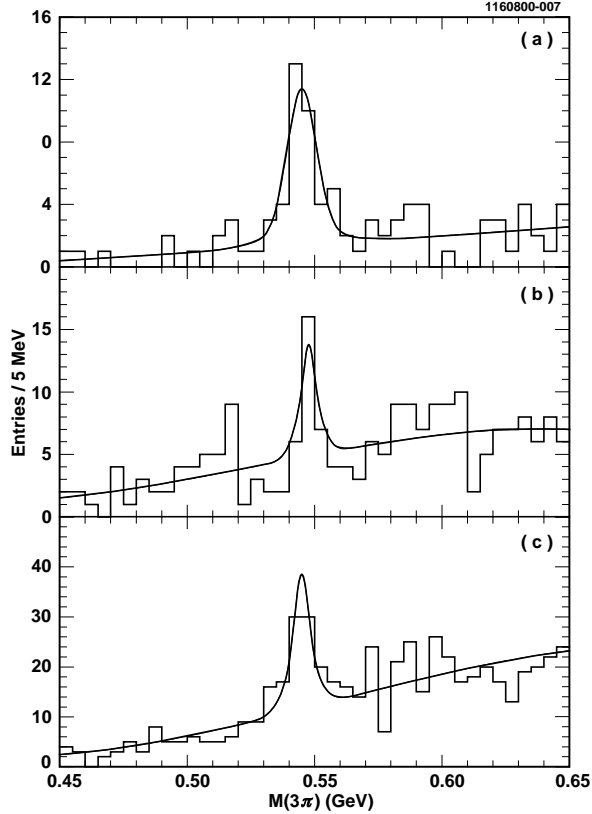, width=8.0cm}}}
\vspace{0.2in}
\caption{Three-pion mass spectra of $\eta$ candidates:
         (a) $M_{3\pi^0}$ in $\tau^-\to 2\pi^- \pi^+ 3\pi^0 \nu_{\tau}$,
         (b) $M_{\pi^+\pi^-\pi^0}$ in $\tau^-\to 2\pi^- \pi^+ 3\pi^0 \nu_{\tau}$,
         (c) $M_{\pi^+\pi^-\pi^0}$ in $\tau^-\to 3\pi^- 2\pi^+ \pi^0 \nu_{\tau}$.
         There are six entries per event in (b) and (c). 
         The solid lines shows fits to the data.} 
\label{figure:eta_res}
\end{figure}

\begin{figure}
\centering
\centerline{\hbox{\psfig{figure=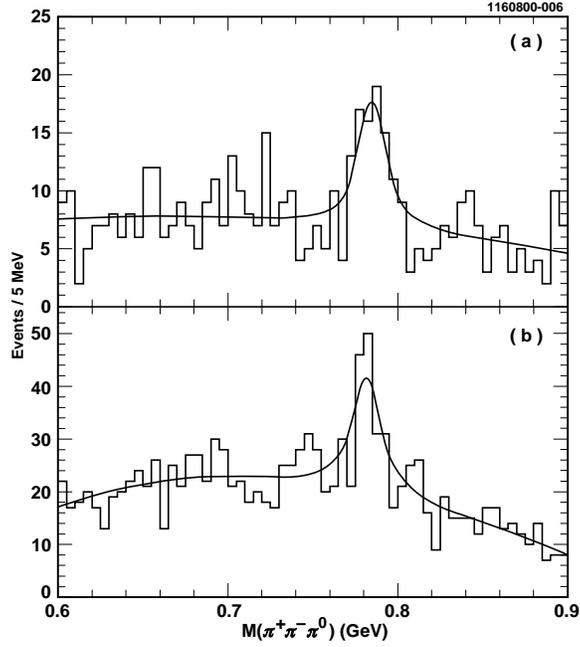, width=8.0cm}}}
\vspace{0.2in}
\caption{Three-pion mass spectra of $\omega$ candidates (six entries per event)
         in (a) $\tau^-\to 2\pi^- \pi^+ 3\pi^0 \nu_{\tau}$
         and (b) $\tau^-\to 3\pi^- 2\pi^+ \pi^0 \nu_{\tau}$. 
         The solid lines shows fits to the data.} 
\label{figure:omega_res}
\vspace{0.2in}
\end{figure}

The six-pion $\tau$ decays can proceed through the $\eta$ or $\omega$
intermediate hadronic states.
Figure~\ref{figure:eta_res} shows the invariant mass spectra of
$3\pi$ combinations in the $\eta$ mass region.
To improve the resolution, the $\pi^0$ candidates have been
kinematically constrained to the nominal $\pi^0$ mass.
In the search for $\eta \to \pi^+\pi^-\pi^0$, we reduce the combinatoric
background by rejecting events with $3\pi^0$ invariant mass within
20~MeV ($\sim$3$\sigma$) of the nominal $\eta$ mass.
There are clear enhancements at the $\eta$ mass in all three spectra.
The invariant mass spectra of $3\pi$ combinations in the $\omega$
mass region are shown in Fig.~\ref{figure:omega_res}.
There is also a clear signal at the $\omega$ mass.
In particular, the decay $\tau^- \to 2\pi^-\pi^+\omega \nu_{\tau}$
is observed for the first time.

The $\eta$ and $\omega$ signals are extracted by maximum likelihood
fits with a dual Gaussian over a polynomial background.
The shapes have been constrained to the Monte Carlo expectations.
The results are summarized in Tables~\ref{table:eta_summary}
and~\ref{table:omega_summary}.

\begin{table}
\caption{Summary of the results for the intermediate states with
an $\eta$ meson.
All errors are statistical, except the second errors in the branching
fractions, which are systematic.}

\begin{center}
\begin{tabular}{lccc} 
Decay mode & $2\pi^-\pi^+\eta \nu_{\tau}$ & $\pi^-2\pi^0\eta\nu_{\tau}$ 
                                          & $2\pi^-\pi^+\eta \nu_{\tau}$\\

      & $\eta \to 3\pi^0$ & $\eta \to \pi^+\pi^-\pi^0$& $\eta \to \pi^+\pi^-\pi^0$\\
\hline
Data (events)          &$32.1 \pm 6.7$  & $15.4 \pm 5.4$       &  $52.4 \pm 10.4$ \\
$q \bar q$ bg (events) & $1.9 \pm 1.5$  & $0.2^{+1.0}_{-0.2}$  &  $5.2  \pm  3.2$ \\
$\tau$ bg (events)     &$0.9 \pm 0.9$   & $2.3 \pm 1.6$        &  $2.5  \pm  2.2$ \\
Efficiency (\%)        &$1.28 \pm 0.05$ & $1.48 \pm 0.05$      &  $4.18 \pm 0.08$ \\
\hline
${\cal B}~(10^{-4})$&$2.9 \pm 0.7 \pm 0.5$& $1.5 \pm 0.6 \pm 0.3$&$1.9 \pm 0.4 \pm 0.3$\\
\end{tabular}
\end{center}
\label{table:eta_summary}
\end{table}

\begin{table}
\caption{Summary of the results for the intermediate states with
an $\omega$ meson. 
All errors are statistical, except the second errors in the branching
fractions, which are systematic.}

\begin{center}
\begin{tabular}{lcc}
Decay mode &    $\pi^-2\pi^0\omega\nu_{\tau}$  &$2\pi^-\pi^+\omega\nu_{\tau}$   \\
\hline 
Data (events)          & $54.3 \pm 11.4$     &  $110.1 \pm 18.6$     \\
$q \bar q$ bg (events) & $1.0^{+2.4}_{-1.0}$ &  $3.5^{+4.2}_{-3.5}$     \\
$\tau$ bg (events)     & $8.7 \pm 3.4$       &  $2.3^{+3.7}_{-2.3}$  \\
Efficiency (\%)        & $1.39 \pm 0.06$     &  $4.06 \pm 0.14$      \\
\hline
${\cal B}~(10^{-4})$ & $1.5 \pm 0.4 \pm 0.3$ &  $1.2 \pm 0.2 \pm 0.1$ \\
\end{tabular}
\end{center}
\label{table:omega_summary}
\end{table}

There are several sources of systematic errors. These include
the uncertainty in the number of produced $\tau^+\tau^-$ pairs (1.4\%),
charged track reconstruction (1\% per track), $\pi^0$ reconstruction
(4\% per $\pi^0$), efficiency (1-5\%) and $\tau$ migration background (4-12\%)
estimates due to limited Monte Carlo statistics, hadronic background
estimates (4-9\%), mixtures of various intermediate states
(1-6\%) and branching fractions of $\eta \to 3\pi^0$ and $\pi^+\pi^-\pi^0$
and $\omega \to \pi^+\pi^-\pi^0$~\cite{PDG98}.
For the extraction of $\omega$ and $\eta$ signals, there are also
systematic errors (5-10\%) resulting from the choice
of combinatorial background shape and fit region.
The branching fractions with systematic errors are listed in
Tables~\ref{table:6pi_summary}-\ref{table:omega_summary}.
The results represent significant improvement in precision
over previous measurements~\cite{PDG98}.
The branching fractions for the two decays with
$\omega$ in the final states are somewhat smaller than the  
recent calculations by Gao and Li~\cite{GaoAndLi}.

The results on the six-pion decays can be compared with the isospin
symmetry and CVC predictions, after correcting for the contributions
from the axial-vector current $\tau^- \to (3\pi)^-\eta \nu_{\tau}$,
which also violates isospin conservation with the decays
$\eta \to 3\pi^0$ and $\pi^+\pi^-\pi^0$.
To reduce the uncertainty in the corrections, we use measurements
from these decays and the decay $\eta \to \gamma\gamma$~\cite{3pieta}
to obtain the average branching fractions:
${\bar {\cal B}}(\tau^-\to 2\pi^-\pi^+\eta\nu_{\tau})=(2.4\pm0.5)\times 10^{-4}$
and
${\bar {\cal B}}(\tau^-\to \pi^-2\pi^0\eta\nu_{\tau})=(1.5\pm0.5)\times 10^{-4}$.
This yields the vector current branching fractions:
${\cal B}_V(\tau^-\to 2\pi^-\pi^+3\pi^0\nu_{\tau})=(1.1 \pm 0.4)\times 10^{-4}$ and 
${\cal B}_V(\tau^-\to 3\pi^-2\pi^+\pi^0\nu_{\tau})=(1.1 \pm 0.2)\times 10^{-4}$,
corresponding to $\sim$50\% and $\sim$65\%, respectively, of the inclusive
six-pion branching fractions.

\begin{figure}
\centering
\centerline{\hbox{\psfig{figure=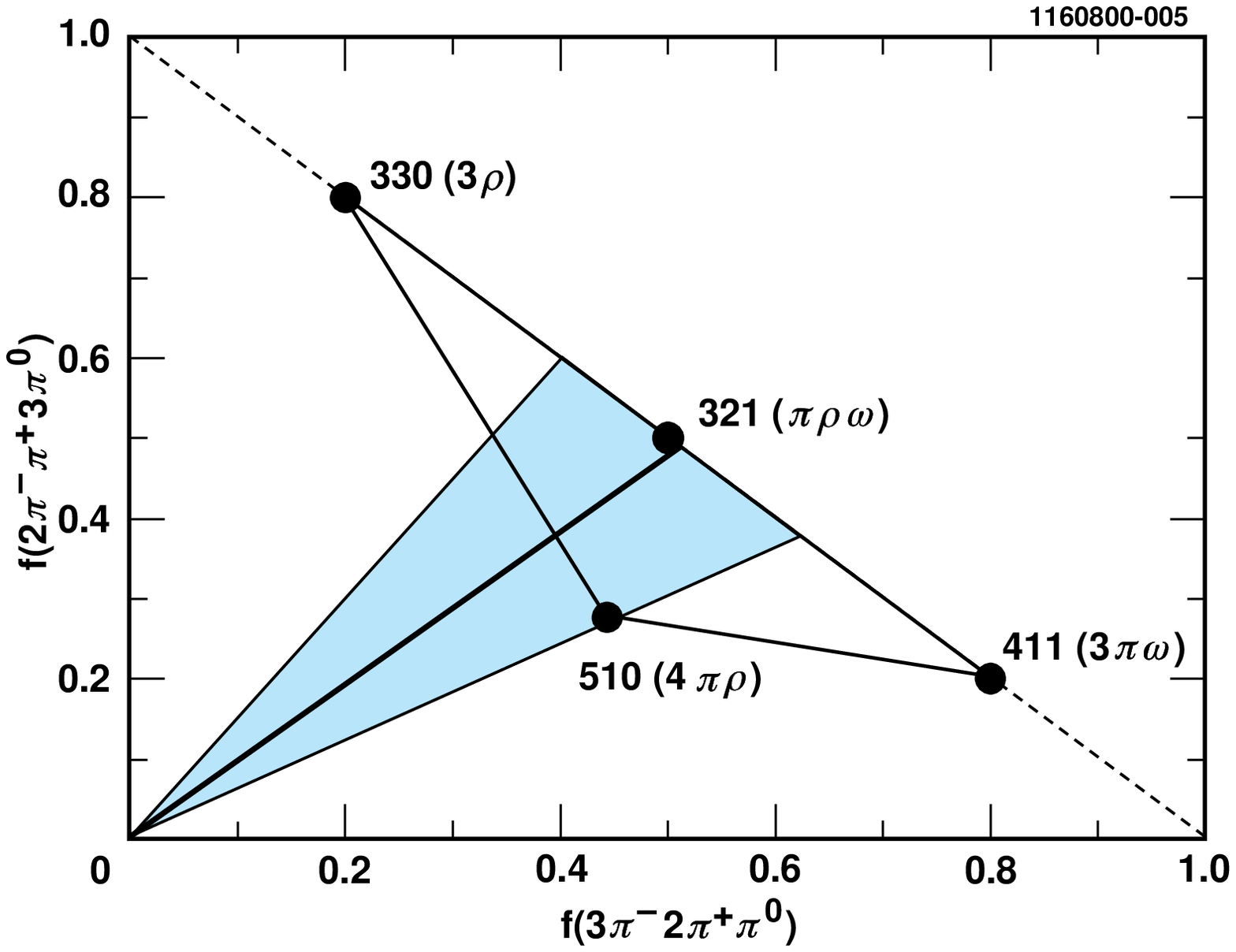, width=8.0cm}}}
\vspace{0.2in}
\caption{Decay fractions of $\tau^-\to (6\pi)^-\nu_{\tau}$.
         The thick solid line through the origin represents the measurement.
         The shaded area indicates the one standard deviation region,
         calculated with correlated errors taken into account.
         The triangle bounded by the dots shows the isospin expectation.}
         
\label{figure:isospin}
\end{figure}

The isospin model~\cite{Pais} classifies $n$-pion final states into
orthogonal isospin partitions and determines their contributions 
to the branching fractions.
The partitions are labeled by three integers $(n_1,n_2,n_3)$,
where $n_3$ is the number of isoscalar subsystems of three pions,
$n_2-n_3$ is the number of isovector systems of two pions, and $n_1-n_2$
is the number of single pions. For  $n=6$ there are four partitions:
510 ($4\pi\rho$), 330 ($3\rho$), 411 ($3\pi\omega$), and 321
($\pi\rho\omega$), denoted according to the lowest mass states.
The isospin model imposes constraints on the relative branching
fractions, which can be tested by comparing the fraction
\begin{eqnarray}
\nonumber
{\rm f}_{2\pi^-\pi^+3\pi^0} = \frac{{\cal B}_V(\tau^-\to 2\pi^-\pi^+3\pi^0\nu_\tau)}
                     {{\cal B}_V(\tau^-\to (6\pi)^-\nu_\tau)}
\end{eqnarray}
to a similar fraction ${\rm f}_{3\pi^-2\pi^+\pi^0}$,
where ${\cal B}_V(\tau^-\to (6\pi)^-\nu_\tau)$
is the sum of the branching fractions for the three six-pion vector decays.
Figure~\ref{figure:isospin} shows ${\rm f}_{2\pi^-\pi^+3\pi^0}$
vs.~${\rm f}_{3\pi^-2\pi^+\pi^0}$ with our new measurement of
the branching fractions.
The measurement is presented as a line because
${\cal B}_V(\tau^-\to \pi^-5\pi^0\nu_\tau)$ has not been measured yet.
The result is consistent with the isospin expectation since
the experimental measurement overlaps with the isospin triangle.
Our result indicates the 321
($\pi\rho\omega$) partition is dominant because the decays
$\tau^-\to \pi^-2\pi^0\omega\nu_{\tau}$ and     
$\tau^-\to 2\pi^-\pi^+\omega\nu_{\tau}$ saturate the
six-pion (vector) decays.

The CVC hypothesis combined with isospin symmetry predicts that
the branching fractions for both 
$\tau^-\to 2\pi^-\pi^+3\pi^0\nu_{\tau}$
and $\tau^-\to 3\pi^-2\pi^+\pi^0\nu_{\tau}$ are greater than
$(2.5 \pm 0.4)\times 10^{-4}$~\cite{Eidelman97}. This is 
somewhat larger than the measured branching fractions
for the six-pion vector decays.
The discrepancy is even more significant if we compare the
predicted inclusive branching fraction
${\cal B}_{CVC}(\tau^-\to (6\pi)^-\nu_{\tau}) = (12.3 \pm 1.9)\times 10^{-4}$
with the sum of the measured six-pion vector
branching fractions under the assumption that
${\cal B}_V(\tau^-\to \pi^-5\pi^0\nu_{\tau})$ is comparable with
or smaller than ${\cal B}_V(\tau^-\to 2\pi^-\pi^+3\pi^0\nu_{\tau})$
and ${\cal B}_V(\tau^-\to 3\pi^-2\pi^+\pi^0\nu_{\tau})$~\cite{isospin}.
This assumption is consistent with the observation that the
six-pion vector decays are saturated by intermediate
states with an $\omega$ meson, which implies a small decay
width for the 510 ($4\pi\rho$) state, the only state that
contributes to the decay $\tau^-\to \pi^-5\pi^0\nu_{\tau}$.
The discrepancy might be explained by sizable presence
of $I=0$ states in the $e^+e^-$ annihilation data that should be 
subtracted before calculating the CVC prediction.

In conclusion, the branching fractions for two six-pion decays
have been measured with much improved precision.
The resonance substructure in the decays has been studied.
In particular, the decay $\tau^- \to 2\pi^-\pi^+\omega \nu_\tau$
has been observed for the first time.
Within the statistical precision,
the decays are saturated by $\eta$ and $\omega$ intermediate states.
The measured branching fractions are in good agreement
with the isospin expectations
but somewhat below the CVC predictions.

We gratefully acknowledge the effort of the CESR staff in providing us with
excellent luminosity and running conditions.
This work was supported by 
the National Science Foundation,
the U.S. Department of Energy,
the Research Corporation,
the Natural Sciences and Engineering Research Council of Canada, 
the A.P. Sloan Foundation, 
the Swiss National Science Foundation, 
the Texas Advanced Research Program,
and the Alexander von Humboldt Stiftung.

\newpage

\begin{thebibliography}{99}
\bibitem{charge} We assume all charged particles in the final states are pions.
                 Charge conjugate states are implied throughout this Letter.
\end{thebibliography}
\end{document}